\documentclass[conference]{IEEEtran}
\IEEEoverridecommandlockouts

\usepackage{cite}
\usepackage{amsmath,amssymb,amsfonts}
\usepackage{algorithmic}
\usepackage{graphicx}
\usepackage{textcomp}
\usepackage{xcolor}
\def\BibTeX{{\rm B\kern-.05em{\sc i\kern-.025em b}\kern-.08em
    T\kern-.1667em\lower.7ex\hbox{E}\kern-.125emX}}
    
\usepackage{url}
\usepackage{verbatim}
\usepackage{alltt} % bold verbatim
\usepackage{listings}
\usepackage{subcaption}

\lstdefinestyle{cinlineasm}{
    language=C,
    keywordstyle=\bfseries,
    morekeywords={__asm__, volatile},
    showstringspaces=false
}

\usepackage{acro}
\acsetup{format/first-long=\itshape}
\DeclareAcronym{bdd}{
  short=BDD,
  long=Behavior-Driven Development,
}

\DeclareAcronym{tdd}{
  short=TDD,
  long=Test-Driven Development,
}

\DeclareAcronym{atdd}{
  short=ATDD,
  long=Acceptance Test-Driven Development,
}

\DeclareAcronym{bdx}{
  short=BDX,
  long=Behavior-Driven Explainability,
}

\DeclareAcronym{xai}{
  short=XAI,
  long=Explainable Artifical Intelligence,
}

\DeclareAcronym{xhw}{
  short=XHW,
  long=Explainable Hardware,
}

\DeclareAcronym{cps}{
  short=CPS,
  long=Cyber-Physical System,
}

\DeclareAcronym{sdlc}{
  short=SDLC,
  long=System Development Life Cycle,
}

\DeclareAcronym{isa}{
  short=ISA,
  long=Instruction Set Architecture,
}

\DeclareAcronym{amo}{
  short=AMO,
  long=Atomic Memory Operation,
}

\DeclareAcronym{vp}{
  short=VP,
  long=Virtual Prototype,
}

\DeclareAcronym{llm}{
  short=LLM,
  long=Large Language Model,
}

\title{Behavior-Driven Explainability
\thanks{Funded by Deutsche Forschungsgemeinschaft (DFG) GRK 2972 "CAUSE" - Project number: 513623283.}}

\author{
\IEEEauthorblockN{Caroline Dominik \hspace{2cm} Rolf Drechsler}
\IEEEauthorblockA{
University of Bremen/DFKI\\
Bremen, Germany\\
\{cardom, drechsler\}@uni-bremen.de}
}

\begin{document}

\maketitle

\begin{abstract}
As system complexity has vastly increased, it has become significantly more challenging for a single person or a team to fully understand all aspects of an entire system.
Particularly, this holds when considering all the different stages of a system's development life cycle, such as, e.g., design or maintenance.
But especially for safety-critical systems it is essential that the final design can be trusted. Because of this, {\rm explainability} is becoming an important requirement for modern systems.

In this paper, we aim to achieve this goal by utilizing \textit{Behavior-Driven Development} (BDD), where the expected system behavior is given in the form of structured {\em scenarios}. These scenarios give a sequence of actions for each functionality, and by this can be directly translated into explanations. We introduce this method of deriving explanations based on the specification as \textit{Behavior-Driven Explainability} (BDX). While applicable at any development stage or abstraction level, a case study for the explanation of exceptions in a RISC-V processor shows the support this concept adds during system design.
\end{abstract}


%%%%%%%%%%%%%%%%%%%%%%%%%%%%%%%%%%%%%%%%%%%%%%%%%%%%%%%%%%%%
\section{Introduction}
%%%%%%%%%%%%%%%%%%%%%%%%%%%%%%%%%%%%%%%%%%%%%%%%%%%%%%%%%%%%

System development has long outgrown being only about implementing the system.
Already in 1995, it was, e.g., estimated that 50\% of the project resources were designated for testing activities~\cite{KF1995}.
And since then, the complexity of technology has vastly increased.
Now, technical systems are considered a central component in our daily life.
As this includes safety-critical systems, such as airplanes or medical diagnosis tools, the quality requirements for those systems can be significant.
To meet these high demands while still restricting the production costs, different approaches for system development have been established.
Some methods cover the entire development process~\cite{B2010}, while others aim to improve single activities during development, such as \ac{tdd}~\cite{B2003}.
In \ac{tdd}, test cases are written in a first step, so that implementation can be added in a second step with the aim of satisfying these tests.
This way, focus remains on the desired functionality of a system and no unnecessary implementation is added.
Based on \ac{tdd}, \ac{bdd}~\cite{N2006}\cite{WH2012} has emerged as another development approach.
Here, the desired system behavior is defined in more abstract scenarios instead of test cases, but implementation is still only started after the behavior has been specified.

But the desire for system quality cannot be fulfilled by testing alone.
To establish a more fundamental trust, it is necessary to create transparency regarding how the inner mechanisms of a system work.
With this goal in mind, \textit{explainability} has emerged as new non-functional requirement for systems (see, e.g.,~\cite{KBBLOS2019}).
When a question about a system's action is answered by an explanation, a more thorough understanding can be achieved.
This can assist a person when interacting with this system or it can help in identifying errors within the system's implementation.
An explanation can be given to another machine as well to support it in its own decision-making during their deployment.
To produce such an explanation, the relevant system information has to be identified and represented in a concise manner.
But since self-explainability in systems has only recently emerged as a research direction, so far the approaches remain at a high level.
They often focus on explaining only specific aspects of a system or the directions on how any explanation can be computed remain vague and include no clear steps for implementation.
In~\cite{CauseDLFG2018} or~\cite{CauseBGGKSSVW2019}, e.g., design methodologies for self-explanations are given for embedded and \acp{cps}, but the proposed requirements for a design flow are formulated in an abstract manner, as discussed in more detail later.

To address this lack in more general and also actionable explainability approaches, \ac{bdx} is introduced in this paper.
Based on the concept of \ac{bdd}, a system specification that describes the desired behavior in the form of scenarios is used to derive explanations.
The steps within such a scenario form a chain of actions that defines a certain aspect of the system's behavior.
If an explanation is desired for a certain system event, it can then be derived by identifying the relevant scenarios and reformulating the resulting chain of actions.
It is demonstrated in this paper, how \ac{bdx} can support various activities during the development process and how it can be applied in different types of situations which require an explanation.
To give direction on how it can be implemented, the approach is formalized and steps for computation are identified and illustrated with examples.

The paper is structured as follows.
First, an introduction to self-explainability is given in Section~\ref{sec_explainability} by outlining the origin of this concept and its current research directions.
Then, \ac{bdd} is described in more detail in Section~\ref{sec_bdd}, so that its extension \ac{bdx} can be established in Section~\ref{sec_bdx}.
This definition is applied to a case study in Section~\ref{sec_casestudy} and the paper is summarized in Section~\ref{sec_conclusion}.

%%%%%%%%%%%%%%%%%%%%%%%%%%%%%%%%%%%%%%%%%%%%%%%%%%%%%%%%%%%%
\section{Self-Explainability}
\label{sec_explainability}
%%%%%%%%%%%%%%%%%%%%%%%%%%%%%%%%%%%%%%%%%%%%%%%%%%%%%%%%%%%%

When studying explainability, it is first necessary to determine what an explanation is and where these definitions come from.
This is done in the following.
Further, the existing approaches for adding explainability to technical systems have to be investigated, for which an overview of application domains and explainability frameworks is given with a special focus on the framework proposed in~\cite{CauseFFD2022}.

%%%%%%%%%%%%%%%%%%%%%%%%%%%%%%
\subsection{Explainability within Philosophy}
%%%%%%%%%%%%%%%%%%%%%%%%%%%%%%

The concept of explainability is closely related to the theory of causation, which has its origins in philosophy.
There, something is a cause $c$ of an effect $e$, if without $c$, $e$ would never have occurred. 
This definition based on \textit{counterfactual dependence} was first mentioned by Hume in Section VII of~\cite{H1900} and later formalized by Lewis in~\cite{L1973}.

These investigations on causality were a necessary foundation for studying explainabilty.
In~\cite{HP2005}, it was defined that the aim of causality is to determine which of some given facts are the causes of an effect, while the aim of an explanation is to provide the necessary information for establishing causality.
For this, an explanation contains events that occurred and causal laws, so that the effect can be derived if the causal laws are applied to the events.
But similar to the definition of causality based on counterfactual dependence, a \textit{counterfactual explanation} is possible as well~\cite{L1990}.
As opposed to complete explanations, the explanation is then given by providing a contrastive event $d$ so that effect $e$ would not have occurred if $d$ had happened instead of cause $c$.

%%%%%%%%%%%%%%%%%%%%%%%%%%%%%%
\subsection{Explainability of Technical Systems}
\label{sec_explainability_systems}
%%%%%%%%%%%%%%%%%%%%%%%%%%%%%%

Within recent years, the concept of explainability has gained attention within computer science.
As digital systems have become a central part of our lives, we have to interact with them frequently and we depend on their correct behavior.
To accept this, a high level of \textit{trust} is required, especially when those systems are used within, e.g., the power grid, the automotive industry or medical equipment.
When establishing trust, gaining an understanding of the system's behavioral mechanisms is more effective than only observing the system's behavior or blindly relying on a formal proof result or the statement of an expert.
This holds especially if the understanding in based on an explanation that is not given by a third party but by the system itself, which is called \textit{self-explanation}.

So far, the most prominent application area of explainability is \ac{xai}~\cite{MWLN2022}, as for many machine learning algorithms there is a trade-off between the quality of the result and the transparency on how the result was computed.
It is noteworthy, that within \ac{xai} another differentiation for the type of explanation than complete and counterfactual is used.
\textit{Global} explanations aim at explaining general events within the system behavior, while \textit{local} explanations aim at explaining specific events.
But various aspects of explainability are studied as well in domains beyond \ac{xai}.
The quality framework proposed in~\cite{CKHS2022} brings explainability requirements into focus w.r.t.~software.
For this, methods are provided for how those requirements can be determined, how fitting methods for explanation can be chosen and how the resulting explanations can be evaluated.
Explainability for software systems is considered in~\cite{SKV2021} as well, but there different situations that require an explanation are identified and classified as explanation cases.
There, the concept of global and local explanations appears as well but with a different naming, when situations are classified as generic of case-specific.
In~\cite{SSBZBP2024}, a framework for \ac{xhw} is given, by analyzing the relevant stakeholders, the requirements for explanations, such as safety or debuggability, and the possible explanation approaches, such as standardizing chip requirements or applying verification.
The approaches for adding self-explanation capabilities to \acp{cps} are more general.
The framework of~\cite{CauseDLFG2018} provides a design methodology for adding self-explanations at runtime to a system.
A self-explanation layer is added including an abstract model representing the system, which is used to derive cause-effect chains.
The MAB-EX framework proposed in~\cite{CauseBGGKSSVW2019} has a similar aim and consists of monitoring and analyzing the system's behavior in a continuous loop to build and transfer explanations when necessary.
The framework in~\cite{CauseFFD2022} aims at defining a general conceptual view for self-explanation in \acp{cps}, regardless of the abstraction level or explanation purpose.
It is described in more detail later.

While this is an incomplete list of the ongoing research on self-explainability in technical systems, it shows its variety.
Different abstraction levels such as software, hardware, or specific system types as well as computational algorithms such as machine learning are considered.
Further, many different research directions regarding explainability are investigated, such as requirements, design or general terminology.

%%%%%%%%%%%%%%%%%%%%%%%%%%%%%%
\subsection{Conceptual Framework for Self-Explanation}
%%%%%%%%%%%%%%%%%%%%%%%%%%%%%%

\begin{figure}
    \centering
    \includegraphics[width=0.95\linewidth]{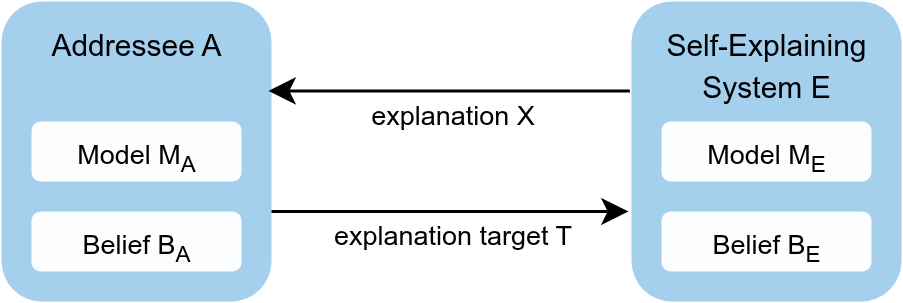}
    \caption{Overview of explanation framework from~\cite{CauseFFD2022}.}
    \label{fig_framework}
\end{figure}

To be able to identify the components of the explanation approach proposed in this paper, the framework of~\cite{CauseFFD2022} is now described more extensively.
It defines a generic explanation pattern $(M,B,T)$, which consists of
\begin{itemize}
\item a \textit{model $M$}, which describes the system's behavior,
\item a \textit{belief $B$}, which denotes the background knowledge about a certain situation,
\item and an \textit{explanation target $T$}, which is the phenomenon of interest.
\end{itemize}
The components and their relation to each other can be seen in Fig.~\ref{fig_framework}
Each component remains abstract so that the framework is applicable in different use cases.
The model $M$, e.g., can be a Kripke structure, an ontology or a Markov chain.

The relevant actors for any explanation are the \textit{self-explaining system $E$} and an \textit{addressee $A$}.
An explanation is required, if the addressee observes a phenomenon $T$, which it cannot explained based on its local model $M_A$ and local belief $B_A$.
This is denoted by $M_A, B_A \models ? T$.
Based on its own local model $M_E$ and local belief $B_E$, the explainer $E$ then returns an \textit{explanation $X$} so that $T$ is justified by local belief for $A$ if $B_A$ is extended by $X$.
This is denoted by 
\begin{equation*}
M_A,B_A + X \models T.
\end{equation*}

Different types of explanations can be defined as well.
For $X$ to be an action-oriented explanation, it has to contain a list of recommended actions, such as specific inputs or a reactive strategy.
For $X$ to be a counterfactual explanation, $X$ has to contradict aspects of belief $B_A$.
The addressee $A$ didn't expect $T$, as it assumes (incorrectly) that the situation is similar to another known situation with a different outcome than $T$.
Then, $X$ overwrites those aspects, denoted by $M_A,B_A \oplus X \models T$.

%%%%%%%%%%%%%%%%%%%%%%%%%%%%%%%%%%%%%%%%%%%%%%%%%%%%%%%%%%%%
\section{Behavior-Driven Development}
\label{sec_bdd}
%%%%%%%%%%%%%%%%%%%%%%%%%%%%%%%%%%%%%%%%%%%%%%%%%%%%%%%%%%%%

The concept of \acl{bdd} was first introduced in~\cite{N2006} as a more intuitive alternative to \ac{tdd}.
Whereas for \ac{tdd}, first tests are written and then the implementation is added so that all tests are passed, for \ac{bdd}, first the behavior of the system is described in the form of scenarios and then the implementation is added accordingly.
While it was originally introduced for the development of software systems, it can be extended to other domains, such as hardware design~\cite{DSGD2012}~\cite{DKSGD2018}, as well.
In the following it is described in more detail, how \ac{bdd} is used for system specification and for automated testing.

%%%%%%%%%%%%%%%%%%%%%%%%%%%%%%
\subsection{Specification in Form of Scenarios}
\label{sec_bdd_specification}
%%%%%%%%%%%%%%%%%%%%%%%%%%%%%%

\begin{figure}
\begin{subfigure}[t]{0.5\textwidth}
\begin{small}
\begin{alltt}
\textbf{Scenario:} Addition of two numbers
  \textbf{Given} Two inputs 3 and 4

  \textbf{When} Applying addition

  \textbf{Then} Result is 7
\end{alltt}
\end{small}
\caption{Scenario.}
\label{fig_scenarioAdditionScenario}
\end{subfigure}%

\vspace*{10px}

\begin{subfigure}[t]{0.5\textwidth}
\begin{small}
\begin{alltt}
\textbf{Scenario Outline:} Addition of two numbers
  \textbf{Given} Two inputs <a> and <b>

  \textbf{When} Applying addition

  \textbf{Then} Result is <a>+<b>

  Examples:
     a |  b | result
   ----|----|----------
     2 |  2 | 4
     1 |  8 | 9
    32 | 15 | 47
\end{alltt}
\end{small}
\caption{Scenario Outline.}
\label{fig_scenarioAdditionOutline}
\end{subfigure}
\caption{Gherkin scenarios specifying addition.}
\label{fig_scenarioAddition}
\end{figure}

A \ac{bdd} scenario is most commonly written using the language Gherkin (see Part I, Chapter 3 of~\cite{WH2012}).
An example scenario specifying addition can seen in Fig.~\ref{fig_scenarioAdditionScenario}.
As highlighted there, each scenario consists of three parts:
\begin{itemize}
    \item \textit{Given} defines the context or current state.
    \item \textit{When} defines the triggering actions occurring in that current state, and can be seen as a transition.
    \item \textit{Then} defines the resulting outcome or next state.
\end{itemize}
The keywords \textit{Given}, \textit{When} and \textit{Then} add structure to each scenario but have no meaning to further implementation, so they can all be replaced by ``*".
Each of those parts consists of several steps, which are concatenated with the keywords \textit{And} and \textit{But}.
Several scenarios combined build a feature, which represents one property of the system's behavior.
Scenarios can be further abstracted into scenario outlines (see Part I, Chapter 5 of\cite{WH2012}) by replacing every specific value by a generic variable written as, e.g., \verb+<var>+.
Then, a table with examples has to be added.
The respective scenario outline for the addition scenario in Fig.~\ref{fig_scenarioAdditionScenario} can be seen in Fig.~\ref{fig_scenarioAdditionOutline}.
In the following, only scenario outlines are used.

If the entire system's behavior is described using Gherkin features, this gives the specification for the system.
During this, the natural language syntax of Gherkin makes the decision on what to have in each scenario more intuitive compared to \ac{tdd}.
Further, it simplifies the communication within the development team.

%%%%%%%%%%%%%%%%%%%%%%%%%%%%%%
\subsection{Automated Testing of Scenarios}
%%%%%%%%%%%%%%%%%%%%%%%%%%%%%%

While one main goal of \ac{bdd} is simplifying the process of communicating about and creating the specification of a system, another is the automated creation of tests.
Each scenario is itself an acceptance test, as it defines the acceptance criteria for how the system is supposed to behave in a certain situation.
Applying these tests to an implementation can be automated by using tools which turn the Gherkin specification into an executable specification. 
The most prominent tool is Cucumber\footnote{\url{https://cucumber.io/} (accessed on July 16th 2026)} which can be applied to several implementation languages, but technology dependent tools such as JBehave\footnote{\url{https://jbehave.org/} (accessed on July 16th 2026)} and JDave\footnote{\url{https://github.com/jdave/JDave} (accessed on July 16th 2026)} for Java or Behave\footnote{\url{https://github.com/behave/behave} (accessed on July 16th 2026)} for Python exist as well.
The Gherkin scenarios are connected to the implementation via step definitions (see Part I, Chapter 2 and 4 of~\cite{WH2012}).
A step definition performs as an interface by calling the respective functions of the system as defined in the \textit{Given} and \textit{When} steps and by comparing the result with the expected behavior as defined in the \textit{Then} steps.
The tests are then automatically executed by the tool.

With this, a Gherkin specification can be used for \ac{atdd} by only adding code in order to increase the number of passed Gherkin tests.
This concept is expanded in~\cite{AHOS2026} to be applicable to \ac{tdd} as well.
For this, the so far introduced Gherkin scenarios which specify the desired behavior of the system are called business scenarios and function as acceptance tests.
But they further define technical Gherkin scenarios, which define how this desired behavior should be implemented and therefore function as unit tests.
In their approach, the three development approaches discussed so far then have a different functionality:
\begin{itemize}
    \item \ac{bdd} acts as the coordinating methodology for improved communication during development by using (business and technical) scenarios for specification.
    
    \item \ac{tdd} and \ac{atdd} act as frameworks within this methodology by implementing code w.r.t.~the scenarios.
\end{itemize}

%%%%%%%%%%%%%%%%%%%%%%%%%%%%%%%%%%%%%%%%%%%%%%%%%%%%%%%%%%%%
\section{Behavior-Driven Explainability}
\label{sec_bdx}
%%%%%%%%%%%%%%%%%%%%%%%%%%%%%%%%%%%%%%%%%%%%%%%%%%%%%%%%%%%%

Now, the concept of \acl{bdx} is introduced.
For this, self-explanation capabilities are added to \ac{bdd}.
The basic idea is that, similar to how each Gherkin scenario can be seen as a test, each scenario can be seen as an explanation as well.
Considering, e.g., the scenario for addition in Fig.~\ref{fig_scenarioAdditionOutline}, an explanation target could be:
\begin{quote}
    Why does applying addition to the inputs $5$ and $10$ result in $15$?
\end{quote}
For this, the respective explanation is then derived from the scenario:
\begin{quote}
    Because when applying addition to two inputs $a$ and $b$, the result is computed by $a+b$, and for $a=5$ and $b=10$ this result is then given by $5+10=15$.
\end{quote}

To demonstrate and formalize \ac{bdx} in the following, the desired scope of application for \ac{bdx} is discussed first.
Then, it is described how this scope is acquired by defining when \ac{bdx} explanations are applied during the system development process and how \ac{bdx} explanations are computed.

%%%%%%%%%%%%%%%%%%%%%%%%%%%%%%
\subsection{Applicability Scope}
%%%%%%%%%%%%%%%%%%%%%%%%%%%%%%

Based on the insights given in Section~\ref{sec_explainability_systems} on how versatile the field of explainability in system is, it is desirable for an explanation approach to be as general as possible.

Therefore, the goal for \ac{bdx} is that it can generate self-explanations for a range of systems.
\ac{bdd} is a fitting foundation for this because, as mentioned before, while it was originally used for software development, the application of \ac{bdd} to hardware in~\cite{DSGD2012} and~\cite{DKSGD2018} shows that it can be extended to different domains.

Further, it should be applicable in different situations that need explanation.
In \cite{SKV2021}, situations are categorized by training, interaction with the system, debugging and validation, of which training and validation require global explanations, while the other two require local ones.
Hence, another requirement for \ac{bdx} is that different types of explanations, such as global and local ones, can be computed. 

%%%%%%%%%%%%%%%%%%%%%%%%%%%%%%
\subsection{Explanation Setup}
%%%%%%%%%%%%%%%%%%%%%%%%%%%%%%

\begin{figure}
    \centering
    \includegraphics[width=0.95\linewidth]{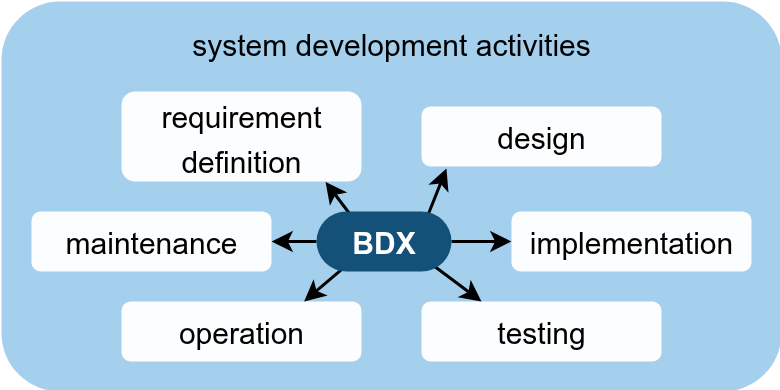}
    \caption{Activities supported by BDX during system development.}
    \label{fig_activities}
\end{figure}

\ac{bdx} can be applied in various ways during system development.
Considering the \ac{sdlc}~\cite{B2010}, the development process consists of the following activities: requirement definition, design, implementation, testing, operation and maintenance.
Depending on the chosen development model, these activities can be traversed in linear progression like, e.g., in the waterfall model or they can be traversed iteratively like, e.g., when using agile methods.
Further, activities can be combined in a single step.
Regardless of which model is chosen though, \ac{bdx} can be applied during any of these activities, as depicted in Fig.~\ref{fig_activities}.
Further, is is not required that each activity is executed according to \ac{bdd}.
The only requirement is that for a behavior of the system to be usable in an explanation, this behavior has to be specified as a Gherkin scenario.
But this Gherkin specification can be incomplete and, further, the implementation does not have to be based on \ac{tdd} or \ac{atdd}.

The main goal of each \ac{sdlc} activity and the value gained by using \ac{bdx} during each of them based on the explanation situations identified in~\cite{SKV2021} is described in the following:
\begin{itemize}
    \item During \textbf{requirement definition}, the system or parts of the system are specified using business scenarios.
    \ac{bdx} can be applied when extending the specification to get a better understanding of the already existing scenarios.
    Further, it can be applied when validating the specification.
    By questioning specific aspects of the specification, the resulting explanation can be evaluated with respect to the desired behavior.
    
    \item During \textbf{design}, the business scenarios are refined by specifying technical scenarios.
    Similar to the requirement definition, \ac{bdx} can be applied to the already existing technical scenarios to extend or to validate them.
    Further, by using the self-explanations of the business scenarios, the process of deriving technical scenarios from them is supported.
    
    \item During \textbf{implementation}, code is written to fulfill the defined specification.
    \ac{bdx} can be applied here to better understand the scenarios when implementing code for them.
    
    \item During \textbf{testing}, it is verified whether the created implementation is correct w.r.t.~the specification.
    Test cases can be derived based on explanations and explanations can support fixing errors detected by failed test cases.
    
    \item During \textbf{operation} the system is used.
    Here, explanations can assist users when they are being trained in using the system.
    They can further support other machines with their own decision-making, e.g., when reconfiguring themselves.
    
    \item During \textbf{maintenance}, the system is modified according to errors that occur during operation or to changing requirements.
    Again, explanations can assist when detecting the cause of an error and when fixing it.
    Further, explanations are useful when the specification is changed or extended by someone without full knowledge of the system's behavior.
\end{itemize}
Depending on the chosen \ac{sdlc} method, the opportunities for and quality of extracting self-explanations can vary.
But regardless of the specific activity, analyzing how \ac{bdx} can be applied throughout the entire development process shows its main gain compared to manually reading the entire specification:
\begin{itemize}
    \item The interaction with the specification is more intuitive.
    \item The reformulation of scenarios as explanations is easier to understand.
\end{itemize}
These perks are most relevant for more complex systems with an extensive specification, as then usually not all parts of the specification are known to all team members.
Further, if there are several causes for the same behavior in question and those causes are at different locations in the specification, manually extracting the aspects of the specification which are relevant in the current situation is a time-consuming and often even impossible task.
Computing the respective \ac{bdx} explanation is faster and less error-prone.

Based on these observations on the situations, where \ac{bdx} can be applied, the framework of~\cite{CauseFFD2022} can be applied.
The resulting components of \ac{bdx} are listed below:
\begin{itemize}
    \item The \textbf{explainer \boldmath$E$} is the self-explaining system.
    \item The \textbf{addressee \boldmath$A$} is a member of the developer team, a user of the system or another machine within the same deployment environment.
    \item The \textbf{model \boldmath$M$} is given by the Gherkin specification, tests and implementation. 
    The explainer's model $M_E$ consists of a complete knowledge of those, while the addressee's model $M_A$ can be an incomplete or even partially incorrect understanding.
    \item The \textbf{belief \boldmath$B$} is some subset of the system's behavior.
    If the behavior in question has implementation and tests, this behavior is the actually observed behavior when executing a test bench.
    Then, the addressee's belief $B_A$ consists of only inputs and outputs, while the explainer's belief $B_E$ additionally contains the internal system states and actions.
    Otherwise, this behavior is a subset of any steps within the Gherkin specification.
    \item The \textbf{explanation target \boldmath$T$} is any set of \textit{Then} steps of the Gherkin specification.
    If the system has been executed, this set is given by the observable system outputs.
    \item The \textbf{explanation \boldmath$X$} is a sequence of actions consisting of \textit{Given} and \textit{When} steps of the Gherkin specification.
    This set can be transformed into natural language to be understandable for the addressee.
\end{itemize}

%%%%%%%%%%%%%%%%%%%%%%%%%%%%%%
\subsection{Explanation Computation}
%%%%%%%%%%%%%%%%%%%%%%%%%%%%%%

Based on the more general observations on when an explanation can be computed during \ac{bdx}, it is now defined how such an explanation can be computed.

For this, $\Sigma$ denotes the set of all possible steps of the considered Gherkin specification, and by that, is the alphabet.
Then, $\sigma \in \Sigma$ is one single step, $s \subseteq \Sigma$ is a scenario and $S = \{s_0, s_1, \dots\}$ the set of all scenarios of the specification.
As already observed in~\cite{DSGD2012}, a Gherkin scenario can easily be expressed as an implication $P \rightarrow Q$.
For this, the \textit{When} steps are the antecedent $P$, the \textit{Then} steps are the consequent $Q$ and the \textit{Given} steps are global assumptions that must hold for the implication.
Using this, a scenario $s \in S$ can be seen as an implication $P_s \rightarrow Q_s$ with $P_s,Q_s \subseteq s$, so that $P_s$ contains the \textit{Given} and \textit{When} steps of $s$ and $Q_s$ contains the \textit{Then} steps of $s$.
Since the keywords \textit{Given}, \textit{When} and \textit{Then} are not clearly defined within the Gherkin language and a distinction of \textit{Given} and \textit{When} is not necessary for explanations, they are summarized as $P_s$ here.
If no keywords are used at all, the steps of $s$ can be separated depending on the context so that the first part gives $P_s$ and the second part gives $Q_s$.
An explanation target $T$ is a tuple $(P_T, Q_T)$ with $P_T \subseteq \Sigma$ restricting the \textit{Given} and \textit{When} steps and $Q_T \subseteq \Sigma, Q_T \neq \emptyset$ restricting the \textit{Then} steps of the behavior that is to be explained.
An explanation $X$ is a set $\{P_{Q_1}, P_{Q_2}, \dots\}$, where each $P_{Q_i}$ denotes a sequence of \textit{Given} and \textit{When} steps.

To compute $X$ for some $T$, the following steps are executed:
\begin{enumerate}
    \item If $T$ is given in natural language, its sets $P_T$ and $Q_T$ are translated so that they use the alphabet $\Sigma$. 
    
    \item Based on $P_T$ and $Q_T$, a set of scenarios $S_T$ is determined, so that for each $s \in S_T$ 
    \begin{itemize}
        \item $Q_T \subseteq Q_s$ holds,
        \item and the overlap of $P_T$ and $P_s$ is maximized.
    \end{itemize}
    Finally, all \textit{Then} steps are removed for each $s \in S_T$, as no \textit{Then} that are additional to the ones of $T$ are of relevance for the explanation. 
    
    \item $X$ is the given, by reformulating the resulting set $S_T$ as a natural language explanation.
    For this, the explanation target is repeated based on $T$ and the cause is given based on $S_T$.
    If possible, elements of $S_T$ can be reduced or summarized according to $T$.
\end{enumerate}

\begin{figure}
    \centering
    \includegraphics[width=0.95\linewidth]{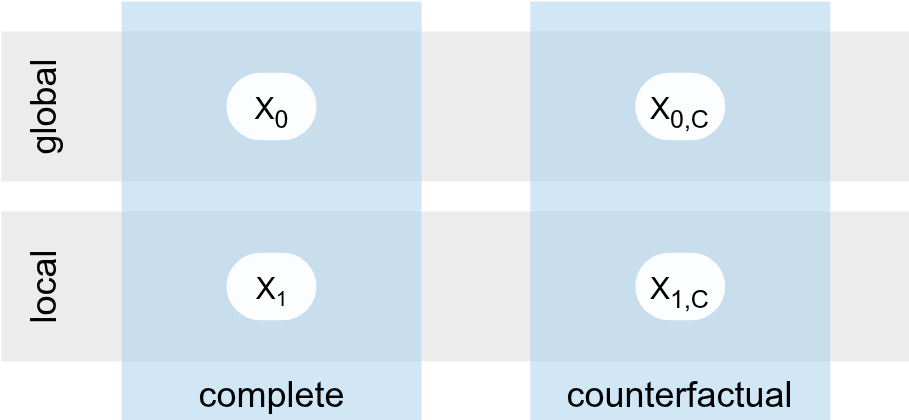}
    \caption{Explanation types considered by BDX.}
    \label{fig_explTypes}
\end{figure}

These steps can be adapted to obtain the different types of explanations introduced in Section~\ref{sec_explainability}, so that \ac{bdx} can compute any explanation within the dimensions given by global vs.~local and complete vs.~counterfactual as depicted in Fig.~\ref{fig_explTypes}.
This is done in the following.

%%%%%%%%%%%%%%%%%%%%%%%%%%%%%%
\subsubsection{Global and Local Explanations}
%%%%%%%%%%%%%%%%%%%%%%%%%%%%%%

A global explanation considers a general aspect of the system's behavior, so here the explanation target does not restrict the context, meaning $P_T = \emptyset$ holds.
A local explanation on the other hand considers a specific execution detail, so here $P_T \neq \emptyset$ holds.
Then, $X$ contains an explanation of the general behavioral mechanisms in both cases, but for the local explanation the values of the specific behavior that is viewed can be inserted in the steps of the global explanation to connect them with each other.

%%%%%%%%%%%%%%%%%%%%%%%%%%%%%%
\subsubsection{Counterfactual and Action-Oriented Explanations}
%%%%%%%%%%%%%%%%%%%%%%%%%%%%%%

The explanation generated by the three computation steps defined earlier is a complete explanation.
A counterfactual explanation can be added to $X$ by naming a minimal change to $S_T$ so that the events of $T$ don't occur.
Further, an action-oriented explanation can be added to $X$ by giving directions how $S_T$ has to be changed to achieve the counterfactual explanation. 

%%%%%%%%%%%%%%%%%%%%%%%%%%%%%%%%%%%%%%%%%%%%%%%%%%%%%%%%%%%%
\section{Case Study}
\label{sec_casestudy}
%%%%%%%%%%%%%%%%%%%%%%%%%%%%%%%%%%%%%%%%%%%%%%%%%%%%%%%%%%%%

To demonstrate the application of \ac{bdx}, a specific scenario is described in the following.
The case study considers the design of a RISC-V based system, for which a RISC-V \ac{vp} (see, e.g.,~\cite{HGPD2020}) is used.
First, the scenario is described in more detail.
Then, \ac{bdx} is applied for a global explanation target during the design of technical scenarios for the RISC-V \ac{vp}.
Finally, \ac{bdx} is applied for a local explanation target during the implementation of C code that is executed on the RISC-V \ac{vp}.
These use cases cover all types of explanations considered within this paper as shown in Fig.~\ref{fig_explTypes}.

%%%%%%%%%%%%%%%%%%%%%%%%%%%%%%
\subsection{Scenario Description}
%%%%%%%%%%%%%%%%%%%%%%%%%%%%%%

A RV32IA variant of the RISC-V~\cite{RiscvUnpriv20260120}~\cite{RiscvPriv20260120} architecture is used, meaning the 32-bit base instruction set ``I" and, additionally, the atomic instruction extension ``A" are used.
To ensure explainability, the \ac{vp} is specified using Gherkin features and implemented accordingly.
In this case study, specifically the Gherkin feature defining address-misaligned exceptions for store operations and \acp{amo} is considered.
The specification is according to the RISC-V \ac{isa}:

\begin{itemize}
    \item The Unprivileged \ac{isa}~\cite{RiscvUnpriv20260120} defines when exceptions are thrown:
    The store/AMO address misaligned exception can be thrown if store instructions use memory locations that are not naturally aligned.
    RV32IA has a 32-bit address space that is byte-addressed.
    Hence, when storing a 32-bit word with the instruction ``SW" at memory byte $m$, then $m$ has to be evenly divisible by four to be naturally aligned.
    Analogously, when storing a 16-bit half word with the instruction ``SH" at $m$, then $m$ has to be evenly divisible by two.
    The conditional store word instruction ``SC.W" of the ``A" extension behaves similar to the ``SW" instruction.
    Based on the \ac{isa}, a different handling is possible as well, hence the Gherkin feature here decides that an exception is thrown in those cases.
    
    \item The Privileged \ac{isa}~\cite{RiscvPriv20260120} defines how exceptions are internally represented:
    Once the exception is triggered, the control and status registers are modified.
    The \textit{Machine Cause} (mcause) register contains the value ``$6$" to specify the exception type ``Store/AMO address misaligned".
    The \textit{Machine Trap Value} (mtval) register contains the respective memory location of the instruction as additional information for this exception.
\end{itemize}

%%%%%%%%%%%%%%%%%%%%%%%%%%%%%%
\subsection{Global BDX during Design Definition}
%%%%%%%%%%%%%%%%%%%%%%%%%%%%%%

\begin{figure}
\begin{small}
\begin{alltt}
\textbf{Feature F0:} Store/AMO Address Misaligned 
  Exception


\textbf{Scenario Outline F0.0:} SH instruction
  \textbf{When} executing a SH instruction
  And memory location in register <rs1>
  And offset <imm>
  And ((<rs1>+<imm>) % 2) != 0

  \textbf{Then} an exception is thrown
  And register <mcause> = 6
  And register <mtval> = <rs1>+<imm>

  Examples:
    rs1 | imm | exception thrown
   -----|-----|------------------
     0  |  2  |  no
     0  |  4  |  no
     3  |  1  |  no
     5  |  0  |  yes
     8  |  2  |  yes

   
\textbf{Scenario Outline F0.1:} SW instruction
  \textbf{When} executing a SW instruction
  And memory location in register <rs1>
  And offset <imm>
  And ((<rs1>+<imm>) % 4) != 0

  \textbf{Then} an exception is thrown
  And register <mcause> = 6
  And register <mtval> = <rs1>+<imm>

  Examples:
    rs1 | imm | exception thrown
   -----|-----|------------------
     0  |  2  |  yes
     0  |  4  |  no
     3  |  1  |  no
     5  |  0  |  yes
     8  |  2  |  yes


\textbf{Scenario Outline F0.2:} SC.W instruction
  \textbf{When} executing a SC.W instruction
  And memory location in register <rs1>
  And (<rs1> % 4) != 0

  \textbf{Then} an exception is thrown
  And register <mcause> = 6
  And register <mtval> = <rs1>

  Examples:
    rs1 | exception thrown
   -----|------------------
     2  |  yes
     4  |  no
     5  |  yes
     8  |  no
\end{alltt}
\end{small}
\caption{Gherkin feature specifying RISC-V store/AMO address misaligned exceptions.}
\label{fig_featureException}
\end{figure}

For the design definition of the RISC-V \ac{vp} according to \ac{bdd}, the store/AMO address misaligned exception has to be specified using technical scenarios.
According to the RISC-V \ac{isa}, the Gherkin feature in Fig.~\ref{fig_featureException} can now be derived.
The feature and the scenarios are enumerated, which is not part of the Gherkin syntax but done here for referencing them more easily in the following.
By applying \ac{bdx}, a developer can now validate the correct design of those exceptions.
For this, the explanation target $T_0$ in natural language is given by:
\begin{quote}
    When are store/AMO address-misaligned exceptions thrown?
\end{quote}
W.r.t.~the Gherkin specification, this is formalized as
\begin{equation*}
    T_0 = (P_{T_0}, Q_{T_0}) = (\emptyset, \{exception~is~thrown, mcause=6\}).
\end{equation*}
Here, only the \textit{Then} steps are restricted by $Q_{T_0}$.
As $T_0$ does not specify a certain execution, no further restrictions $P_{T_0}$ for the \textit{Given} and \textit{When} steps are defined.
Hence, a global explanation is required.

The respective explanation $X_0$ (compare Fig.~\ref{fig_explTypes}) is derived from the Gherkin specification of the \ac{vp}.
All three scenarios of the feature $F0$ in Fig.~\ref{fig_featureException} apply, because their \textit{Then} steps contain $Q_{T_0}$ and no overlap with their \textit{Given} and \textit{When} steps is possible.
Therefore, $S_0 = \{F0.0, F0.1, F0.2\}$ and $X_0$ is given by a combination of the \textit{Given} and \textit{When} steps of all scenarios of $S_0$:
\begin{quote}
    A store/AMO address-misaligned exception is thrown when instruction ``SH", ``SW" or ``SC.W" is executed and $(a \% b) \neq 0$ holds.
    For ``SH" and ``SW" $a$ is given by the addition of memory location $rs1$ and offset $imm$, while for ``SC.W" $a$ is given by $rs1$.
    For ``SW" and ``SC.W" $b$ is $4$, while for ``SH" $b$ is $2$.
\end{quote}
A counterfactual explanation $X_{0,C}$ (compare Fig.~\ref{fig_explTypes}) can be added by changing $S_0$ so that $T_0$ is avoided.
To keep the change minimal, the operands of the instructions used in $S_0$ are modified before any opcodes are changed.
The result is an action-oriented explanation as well:
\begin{quote}
    To avoid store/AMO address-misaligned exceptions, ensure only values for $a$ are used, for which $(a \% 4) = 0$, or $(a \% 2) = 0$ respectively, holds.
\end{quote}

This application of \ac{bdx} shows how the efficiency of validating a system's design can be increased.
Whereas manual validation would require inspecting every step of every scenario, with \ac{bdx} a condensed explanation can be given for a certain behavior.
Especially when a specific behavior appears in several features, these explanation increase the developer's trust in the design, as no cause of this behavior can be missed.

%%%%%%%%%%%%%%%%%%%%%%%%%%%%%%
\subsection{Local BDX during Implementation}
%%%%%%%%%%%%%%%%%%%%%%%%%%%%%%

\begin{figure}
\begin{lstlisting}[style=cinlineasm]
void f (int a, int b, int m) {
  int result;
  __asm__ volatile (
    "mul %[result], %[a], %[b]\n\t"
    "sw %[result], 0(%[m])\n\t"
    : [result] "=r" (result)
    : [a]"r"(a), [b]"r"(b), [m]"r"(m)
    : "memory"
  );
}
\end{lstlisting}
\caption{C code implementing an example function.}
\label{fig_cExample}
\end{figure}

The Gherkin specification of the RISC-V \ac{vp} further gives the basis for explanations during the design of any C code that is executed on the RISC-V \ac{vp}.
Consider, e.g., the C code snippet in Fig.~\ref{fig_cExample}.
The function $f$ uses in-line assembly to multiply two given integers $a$ and $b$ and store the result at a memory location defined by a given integer $m$.
When executing $f$ with the data stream $[0, 12, 24, 34]$ for $m$, no errors occur during the first three function calls, but an exception is thrown during the fourth.
The resulting explanation target $T_1$ in natural language is: 
\begin{quote}
    Why is an store/AMO address-misaligned exception thrown for $f(a,b,34)$?
\end{quote}
Now, when mapping $T_1$ to the steps in $\Sigma$ this gives $Q_{T_1} = (exception~is~thrown, mcause=6)$ and $P_{T_1}$ is the sequence of internal system actions starting from $f$ being called with $m=34$.
Here, these actions are all software actions executed by the C code and the \ac{vp}.
Compared to the previous explanation target $T_0$, now a specific execution is given via $P_{T_1}$.
Hence, a local explanation is required.

The explanation $X_1$ (compare Fig.~\ref{fig_explTypes}) for $T_1$ is derived by comparing the observed system behavior given by $Q_{T_1}$ and $P_{T_1}$ with the chain of actions defined by the parts of each scenario in the \ac{vp} specification.
Again, the \textit{Then} steps of all scenarios of feature $F0$ in Fig.~\ref{fig_featureException} contain the steps of $Q_{T_1}$.
But the \textit{When} steps of scenario $F0.1$ have the biggest overlap with the execution selected by $P_{T_1}$, because only this scenario contains the step ``When executing a SW instruction".
So $S_1 = \{F0.1\}$ holds.
Hence, by formulating the scenario as an explanation $X_1$, the explanation target $T_1$ is answered:
\begin{quote}
	The store/AMO address-misaligned exception is thrown for $f(a,b,34)$ because this exception is thrown if the condition $(m~\%~4) \neq 0$ holds when executing the instruction ``SW" with memory location $m$.
	And $f(a,b,34)$ contains a ``SW" instruction with $m=34$, for which then $(34~\%~4 = 2)$ and $2 \neq 0$ holds.
\end{quote}
Further, a counterfactual $X_{1,C}$ (compare Fig.~\ref{fig_explTypes}) can be added.
Again, $S_0$ is changed so that $T_0$ is avoided, but now specific values can be used based on the previous executions of the system, for which $T_0$ was not observed: 
\begin{quote}
	In contrast, the store/AMO address-misaligned exception is not thrown for $f(a,b,24)$ because for this function call, the ``SW" instruction is executed with $m=24$ instead, for which then $(24~\%~4 = 0)$ and $0 = 0$ holds.
\end{quote}
For an action-oriented explanation, a specific action can be derived from $X_{1,C}$ as well:
\begin{quote}
	To avoid the store/AMO address-misaligned exception, constrain operand $m$ so that $(m~\%~4) == 0$ always holds.
\end{quote}
For a local explanation to be possible, the system has to be implemented in such a way, that not only the resulting outcome is according to the Gherkin specification, but further the system implementation is close enough to the specification so that for every explanation target, the respective Gherkin scenario can be found.

%%%%%%%%%%%%%%%%%%%%%%%%%%%%%%%%%%%%%%%%%%%%%%%%%%%%%%%%%%%%
\section{Conclusion}
\label{sec_conclusion}
%%%%%%%%%%%%%%%%%%%%%%%%%%%%%%%%%%%%%%%%%%%%%%%%%%%%%%%%%%%%

In this paper, \ac{bdx} was introduced as a novel approach for enabling self-explainability in system development witch a broad scope.
It was analyzed how the approach can be applied throughout the entire system development process and to different types of situations in which explanations are needed.
The possibility to apply \ac{bdx} in such a versatile manner is its main perk.
To illustrate this versatility, the approach was formalized and applied to two different use cases, which covered different activities within the \ac{sdlc} and different explanation types.

This introduction of \ac{bdx} gives way to a variety of further research in that direction, such as the application to more extensive use cases to identify its performance and limitations.
Similar to~\cite{DL2025}, where \acp{llm} are used for deriving \ac{bdd} scenarios from the natural language specification of a circuit, a combination with other technology is worth investigating as well.

\bibliography{refs.bib} 
\bibliographystyle{IEEEtran}

\end{document}